\documentclass[aps,pre,twocolumn,english,showpacs]{revtex4-1}
\usepackage{natbib}
\usepackage[utf8]{inputenc}
\pdfoutput=1
\usepackage[T1]{fontenc}
\usepackage[english, french]{babel} 
\usepackage{amsmath}
\usepackage{amsfonts}
\usepackage{graphicx}
\usepackage{mathtools} 
\usepackage{braket} 
\usepackage[usenames,dvipsnames]{xcolor} 

\newcommand{\lamb}{\ensuremath{\overline{\lambda}}}
\newcommand{\zb}{\overline{z}}

\begin{document}
\title{Nonmonotonic crossover and scaling behaviors in a disordered 1D quasicrystal}

\author{ Anuradha Jagannathan }
\affiliation{Laboratoire de Physique des Solides, CNRS-UMR 8502, Universit\'e
Paris-Sud, 91405 Orsay, France }
\author{Piyush Jeena}
\affiliation{Department of Physics, IIT Mumbai, Powai, Mumbai-400076, India}
\author{Marco Tarzia}
\affiliation{LPTMC, CNRS-UMR 7600, Sorbonne Universit\'e, 4 Pl. Jussieu, F-75005 Paris, France}

\selectlanguage{english}

\date{\today}


\begin{abstract}
We consider a noninteracting disordered 1D quasicrystal in the weak disorder regime. We show that the critical states of the pure model approach strong localization in strikingly different  ways, depending on their renormalization properties. A finite size scaling analysis of the inverse participation ratios of states (IPR) of the quasicrystal shows that they are described by several kinds of scaling functions. While most states show a progressively increasing IPR as a function of the scaling variable, other states exhibit a nonmonotonic ``re-entrant'' behavior wherein the IPR  first decreases, and passes through a minimum, before increasing. This surprising behavior is explained in the framework of perturbation renormalization group treatment, where wavefunctions can be computed analytically as a function of the hopping amplitude ratio and the disorder, however it is not specific to this model. Our results should help to clarify results of recent studies of localization due to random and quasiperiodic potentials.
\end{abstract}
\pacs{}
\maketitle

\section{Introduction}
It is established that, in many quasiperiodic models, electronic states are multifractal or ``critical''  in the absence of disorder~\cite{multifrac1,delyonpetritis,multifracfibo1,multifracfibo2,multifracharper1,multifracharper2}. The consequences of adding disorder to a quasiperiodic Hamiltonian have been studied as well, and rigorous arguments~\cite{delyon,simon} predict that the addition of (uncorrelated short-range) disorder, however weak, will result in all states becoming localized. While this statement holds for infinite systems, in this paper we consider {\it finite} samples (i.e., smaller than the localization length) to understand how critical states of the pure system change under addition of weak disorder. We will show that the answer to this question depends on the renormalization properties of the states, leading to different kinds of scaling functions for this problem.  One motivation for our study comes from the recent experimental~\cite{experiments1,experiments2,experiments3,coldatoms1,coldatoms2,coldatoms3} and theoretical~\cite{dave,alet,MBLlit1,MBLlit2,MBLlit3} work on many-body localization in interacting quasiperiodic systems where the question of differences in the nature of the transition for quasiperiodic (also called pseudorandom) versus random potentials has been raised~\cite{MBLlit1,MBLlit2,MBLlit3}. 

In this paper, we consider the noninteracting model to show that interesting new phenomena can occur when random disorder is added to deterministic but nonperiodic order. 
Considering a tight-binding model on Fibonacci approximant chains in the weak disorder regime where the chain length is much smaller than the putative localization length, we show that a large (but sub-extensive) set of states exhibit a nonmonotonic approach to strong localization. This implies that some states are initially delocalized, in the sense that their inverse participation ratio (IPR) starts to {\it decrease} with disorder.  These states subsequently begin ``relocalizing'' when the disorder exceeds a certain value, as one expects, and as verified in other studies~\cite{liu}. We speculate that this type of nonmonotonic behavior could occur in a generic way when the pure quasicrystal is perturbed. 

The paper is organized as follows: Sec.~\ref{sec:model} introduces the model; Sec.~\ref{sec:fss} describes the finite size scaling analysis of the Inverse Participation Ratio obtained from exact diagonalizations of disordered Fibonacci approximants; Sec.~\ref{sec:rg} discusses the results obtained by perturbative renormalization group (RG) theory, with physical interpretation of the different scaling phenomena present; Finally, Sec.~\ref{sec:conclusions} gives a discussion of the results, along with perspectives.

\section{Hopping model on disordered Fibonacci chains} \label{sec:model}
The model considered here is a tight-binding problem of the following form  
\begin{equation} \label{eq:H}
H =   \sum_i t_i \left( \ket{i} \bra{i+1} + \ket{i+1} \bra{i} \right) \, ,
\end{equation}
In this Hamiltonian, the hopping amplitudes $t_i=t_i^{(0)}+\epsilon_i$ are perturbed from the values $t_i^{(0)}$, the initial ``pure'' system hopping amplitudes which can take two values, $t_A$ or $t_B$ according to the deterministic Fibonacci sequence described below.  The site energies are all assumed to be equal and can be set to zero by defining the origin suitably. The properties of the pure Hamiltonian depend on a {\it single} parameter, namely  the hopping ratio $\rho = t_A/t_B$ henceforward supposed to be in the range $0\leq \rho\leq 1$ (in the following we will set $t_B=1$ without loss of generality). 
As customary in the Anderson localization literature, the random bond perturbations, $\epsilon_i$, are chosen as i.i.d. random variables taken from 
a uniform distribution in the interval $[-W/2,W/2]$, such that $\langle \epsilon\rangle=0$ and $\langle \epsilon^2\rangle = W^2/12$. The value of $W$ thus denotes the disorder strength.
The sequence of hopping amplitudes $t_A$ and $t_B$ in the pure system corresponds to letters A and B of a specific series of chains leading in the infinite size limit to the Fibonacci quasicrystal. These chains  $C_{n}$, termed approximants,
can be built iteratively by concatenation, namely $C_{n+1} = C_{n}C_{n-1}$. With initial conditions $C_0=B$ and $C_1=A$, the next few chains are AB, ABA, ABAAB, and so on. The lengths of these chains then obey the Fibonacci recursion relation $L_{n+1} = L_{n}+L_{n-1}$ with initial conditions $L_0=L_1=1$, and the ratio of lengths of successive chains $L_{n}/L_{n+1}$ tends to the (inverse) golden mean $\omega=(\sqrt{5}-1)/2$, in the limit $n\to \infty$. 

The properties of the pure model with no disorder ($\epsilon_j=0 ~\forall j$) have been discussed in many classic papers, using a variety of methods, notably the powerful trace map method~\cite{multifracfibo1,multifracfibo2}. It is known that all states are delocalized in the sense of a vanishing Lyapunov exponent~\cite{delyonpetritis}, but critical. Detailed information on spectrum and states have been obtained using the perturbative RG introduced by Niu and Nori and Kalugin et al.~\cite{niunoriPRL,niunoriPRB,kalugin}. This approach gives quantitatively good predictions for the spectrum, eigenstates, and the quantum diffusion properties of wave packets for $\rho \ll 1$~\cite{pertRG1,pertRG2, thiemschreiberJPCM2011,macePRB2016}. 
We will extend this approach to the disordered case and use it to interpret our numerical results.

\begin{figure}[h]
\centering
 \includegraphics[width=0.48\textwidth]{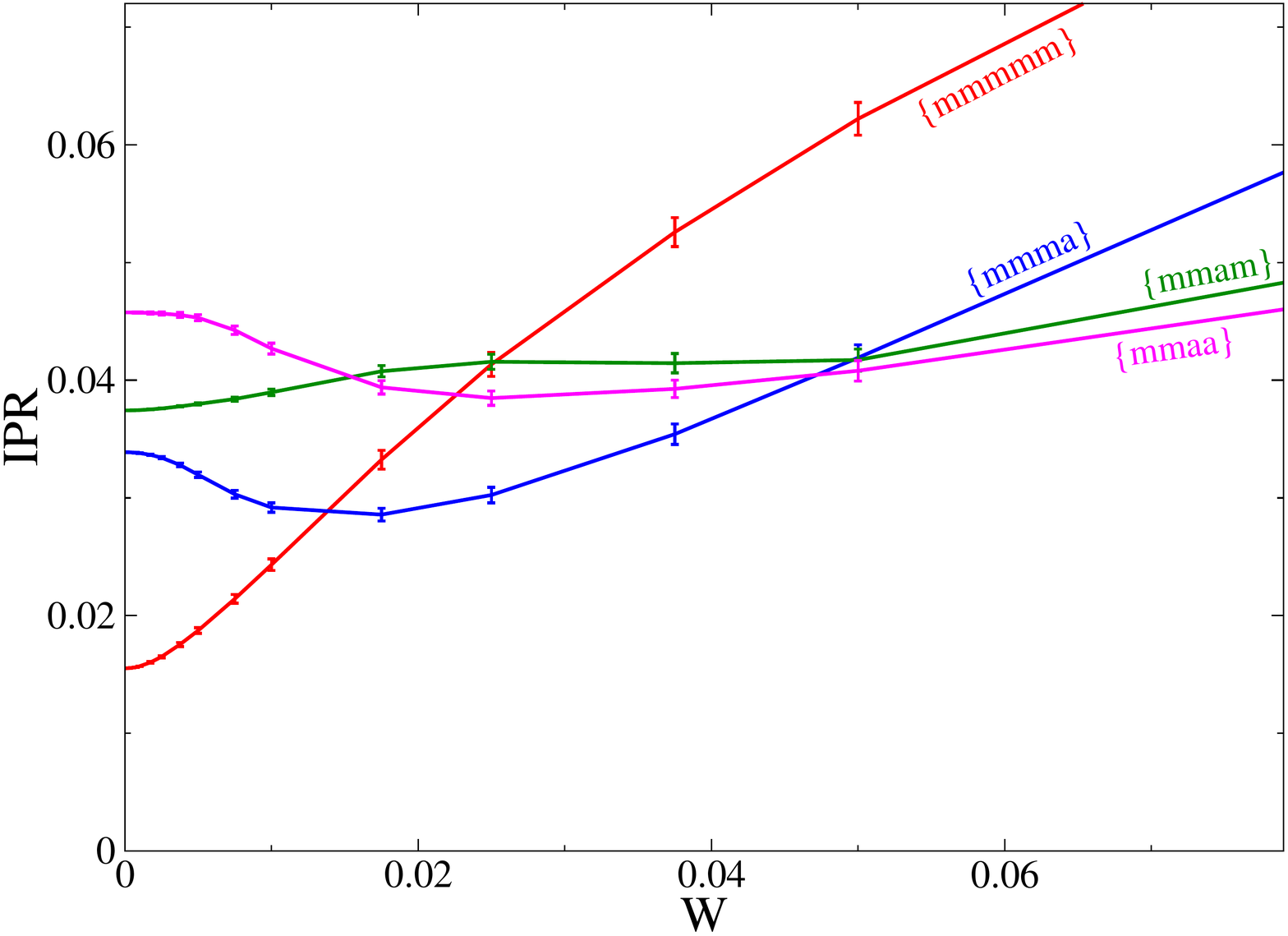} 
 \vspace{-0.4cm}
\caption{Average IPR 
versus $W$ for four states ($\alpha = 1$, red, $\alpha=3$, blue, $\alpha=7$, magenta, and $\alpha=6$, green)on the $n=10$ disordered chain ($L=89$ and $t_A/t_B=1/2$) obtained from exact diagonalizations (averages are performed over $262144$ independent disorder realizations).  Level indices are indicated as are the RG paths (see text). The typical size of the IPR fluctuations is shown.  
}
\vspace{-0.1cm}
\label{dataIPR.fig}
\end{figure}

\section{Disorder dependence and finite size scaling of the IPR} \label{sec:fss}
The averaged IPR corresponding to a given (normalized) eigenstate $\alpha$ ($\alpha=1,...L$) as a function of the disorder strength $W$ is defined by
\begin{eqnarray}
\mathcal{I}_\alpha(W,L) = \left \langle \sum_{i=1}^L \vert \psi_\alpha(i)\vert^4 \right \rangle \, ,
\end{eqnarray}
where the brackets stand for the average over disorder. Throughout this paper we will label the states $\vert \alpha \rangle$ according to their increasing energies, such as $E_1 < E_2 < \ldots < E_L$, and we compare characteristics of states of given $\alpha$ for different system sizes,  as fixed $\alpha$ corresponds to states of given RG path (as described in the next section).  $\mathcal{I}$ is just one of the set of $q$-moments of the probability of presence on each site. We consider the IPR ($q=2$) in this paper, as an indicator of localization adequate for our noninteracting model (for interacting case see~\cite{varmaPRB} for a discussion of the diagnostic tool involving the Kohn localization tensor). Recall that for large system size $L$, $\mathcal{I} \rightarrow L^{-D_2}$ with $D_2$ having the value 1 for an extended state, 0 for a localized state and a value in-between for a critical state. In the Fibonacci chain approximants,  the pure system IPR values $\mathcal{I}_\alpha(0,L_n)$  fluctuate irregularly in a self-similar fractal way with the index $\alpha$. 
In the presence of disorder, the IPR evolve as illustrated in Fig.~\ref{dataIPR.fig} which shows $\mathcal{I}$ as a function of $W$ computed from exact diagonalizations of $n=10$ chains ($89$ sites) for $\rho=0.5$. Four different levels are shown to illustrate the different behaviors which are seen. While the level at the lower band edge ($\alpha=1$) shows a steep increase with $W$,  others (such as $\alpha=3$ and $\alpha=7$) show nonmonotonic behavior. The character strings in the figure are RG paths of each level, detailed in the next section. 

\begin{figure}[h]
\center
\includegraphics[width=0.49\textwidth]{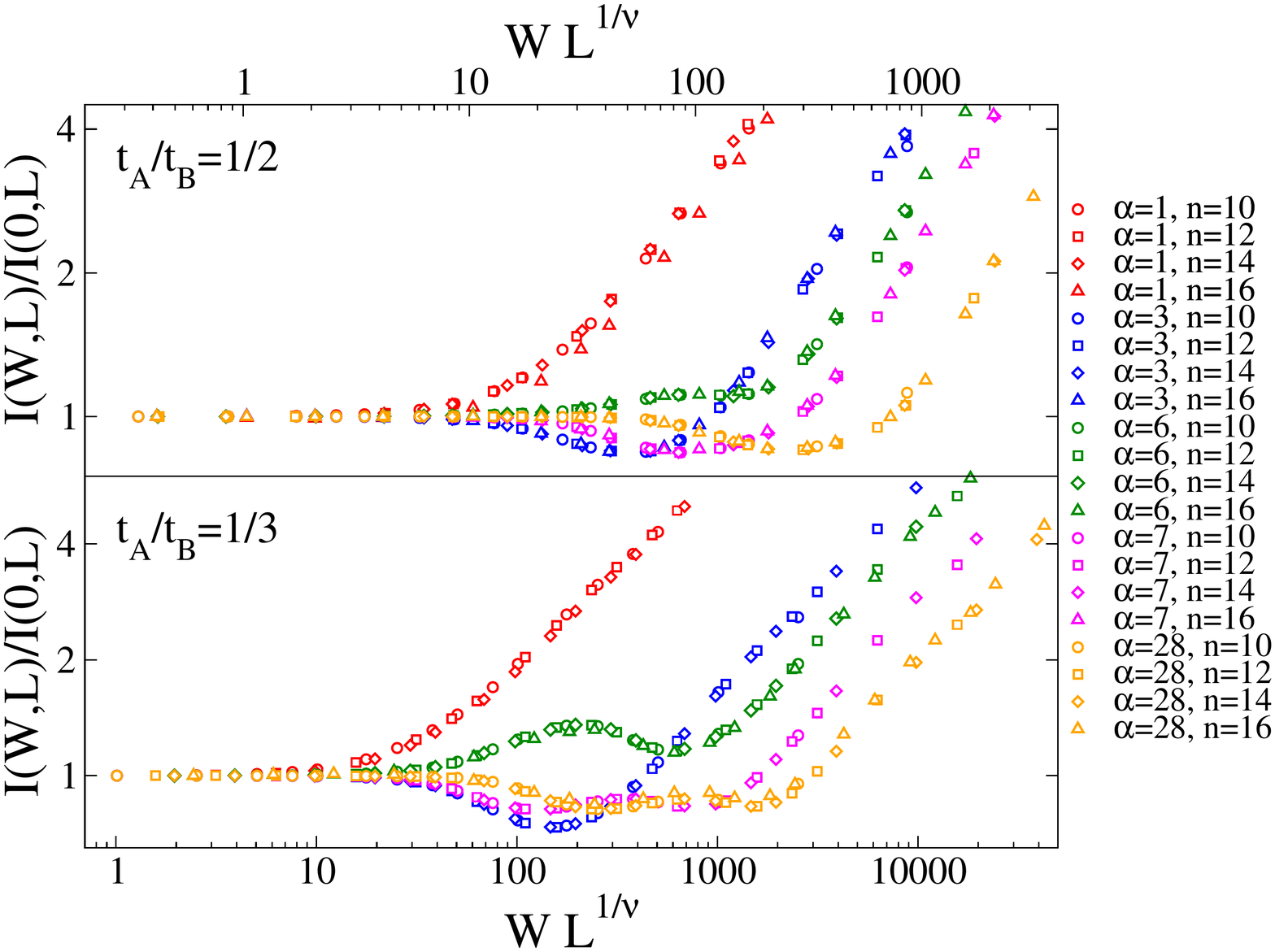}  
\vspace{-0.4cm}
	\caption{Averaged IPR (normalized to ${\cal I}(0,L)$) of several states ($\alpha=1,3,6,7,28$) and system sizes ($n=10,12,14,16$) and for $t_A / t_B=1/2$ (top panel) and $t_A / t_B=1/3$ (bottom panel) obtained from exact diagonalizations, showing data collapse as a function of the scaling variable $W L^{1/\nu}$ with $\nu=1.7$ and $\nu=1.9$ respectively. A similar behavior is found for all individual levels followed from one generation to the next. The nonmonotonicity of the IPR is more pronounced for smaller $\rho$ and disappears continuously in the periodic limit ($\rho \to 1$).}
	\vspace{-0.1cm}
\label{rescaling.fig}
\end{figure}

As in the periodic model the critical point corresponds to $W_c=0$ (any disorder however weak localizes the critical states on the Fibonacci chain) with the localization length given by $\xi \sim W^{-\nu}$, where the $\nu$ is the correlation length exponent. In the weak disorder regime, the IPR is expected to have the scaling form 
\begin{eqnarray}
\frac{\mathcal{I}_\alpha(W,L)}{{\cal I}_\alpha (0,L)} = f_\alpha (L/\xi)
\end{eqnarray}
(see~\cite{roemer} for a more general discussion of finite size effects for the $q$-th moments of wavefunctions near the critical point of the Anderson model). Scaling plots of the IPR obtained from exact diagonalizations for chains of generations $n=10$ to $n=16$ are presented in Fig.~\ref{rescaling.fig} for $\rho = 1/2$ (top panel) and $\rho = 1/3$ (bottom panel), showing good collapse for {\it all} the states when $\mathcal{I}$ is plotted as a function of the scaling variable $W L^{1/\nu}$, for the values $\nu=1.7 \pm 0.04$ and $\nu=1.9 \pm 0.06$ respectively (data are averaged over $262144$, $65536$, $12288$, and $2048$ realizations for $n=10$, $12$, $14$, and $16$ respectively). Several different scaling functions $f_\alpha$ are found, describing the variety of behaviors already seen in Fig.~\ref{dataIPR.fig}. 
Notice that our IPRs are defined with respect to a given state $\alpha$ followed from one generation to the next, implying that the energies of the states move towards
the band edge as the system size is increased.
We have checked numerically that the minima of the IPR of the states displaying the nonmonotonic behavior (as well as the wiggles and secondary minima of the other states) occur essentially when their energies start to cross with those of the neighboring levels (i.e., when the gaps become of the order of the fluctuations of energies), thereby suggesting that the changes of the behavior of the IPR are due to the onset of level repulsion. 

\begin{figure}[h]
\includegraphics[width=0.48\textwidth]{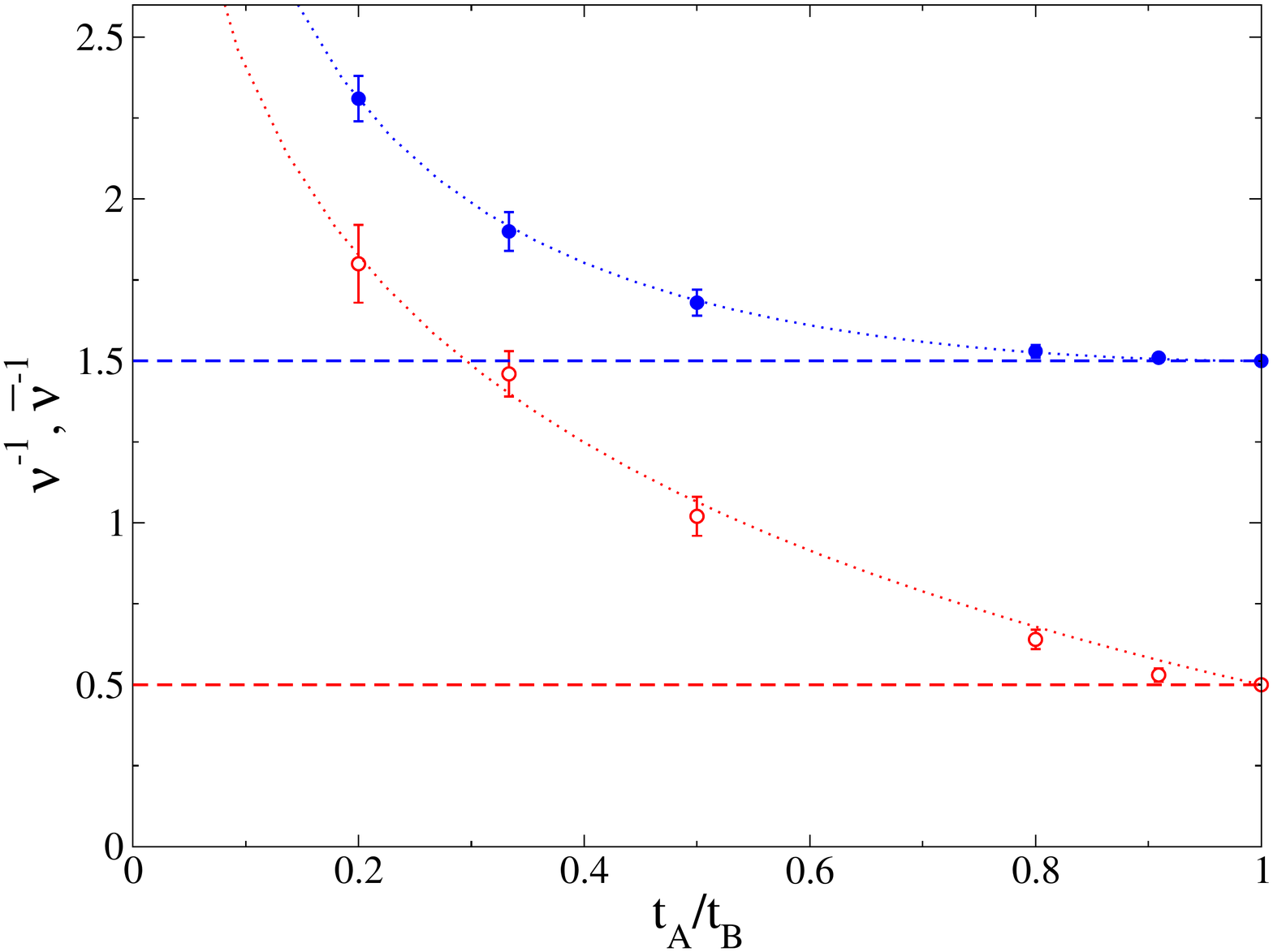}  
\vspace{-0.4cm}
	\caption {Critical exponents $\nu$ and $\overline{\nu}$ 
	as a function of the disorder plotted versus the ratio $t_A / t_B$. The horizontal dashed lines show the standard values of the exponents $\nu=2/3$ and $\overline{\nu}=2$ of the periodic system for the edge and the center states respectively. The dashed lines represent logarithmic fits as $\nu^{-1} \simeq 3/2 + [-0.55 \, \ln \rho]^{1.74}$ and $\overline{\nu}^{-1} \simeq 1/2 -0.82 \, \ln \rho$.}
	\vspace{-0.1cm}
\label{exponent.fig}
\end{figure}

Fig.~\ref{exponent.fig} shows the dependence of the nonuniversal exponent $\nu$ (blue filled circles) on the ratio of hopping amplitudes, $\rho$. The values of $\nu$ descend towards zero as $\rho$ decreases, possibly logarithmically, to zero. However, this limit is difficult to study as some of the gaps between neighboring levels becomes extremely small leading to computational errors. In the periodic limit where $\rho=1$, we find $\nu=2/3$, in agreement with the result obtained for the disorder driven superfluid-insulator phase transition of noninteracting bosons~\cite{continentino}. This value corresponds, in that model as well, to scaling at the band edge. 
As noted by them, $\nu = 2/3$ violates the bound  $\nu\geq 2/d$ established by Chayes et al.~\cite{chayes} for random (interacting) systems, 
as well as the generalized Harris-Luck criterion $\nu\geq 1/d$~\cite{luck} for aperiodic systems. This is not surprising since, unlike the present case, these inequalities apply to transitions at finite disorder. In fact, in contrast with the band edge states, the state in the center of the spectrum at $\langle E \rangle = 0$ ($\alpha = 1 + [L_n/2]$ for $L_n$ odd) scales with the standard universal exponent $\nu=2$ for all hopping ratios $\rho$. 

\begin{figure}[h]
\center
\includegraphics[width=0.48\textwidth]{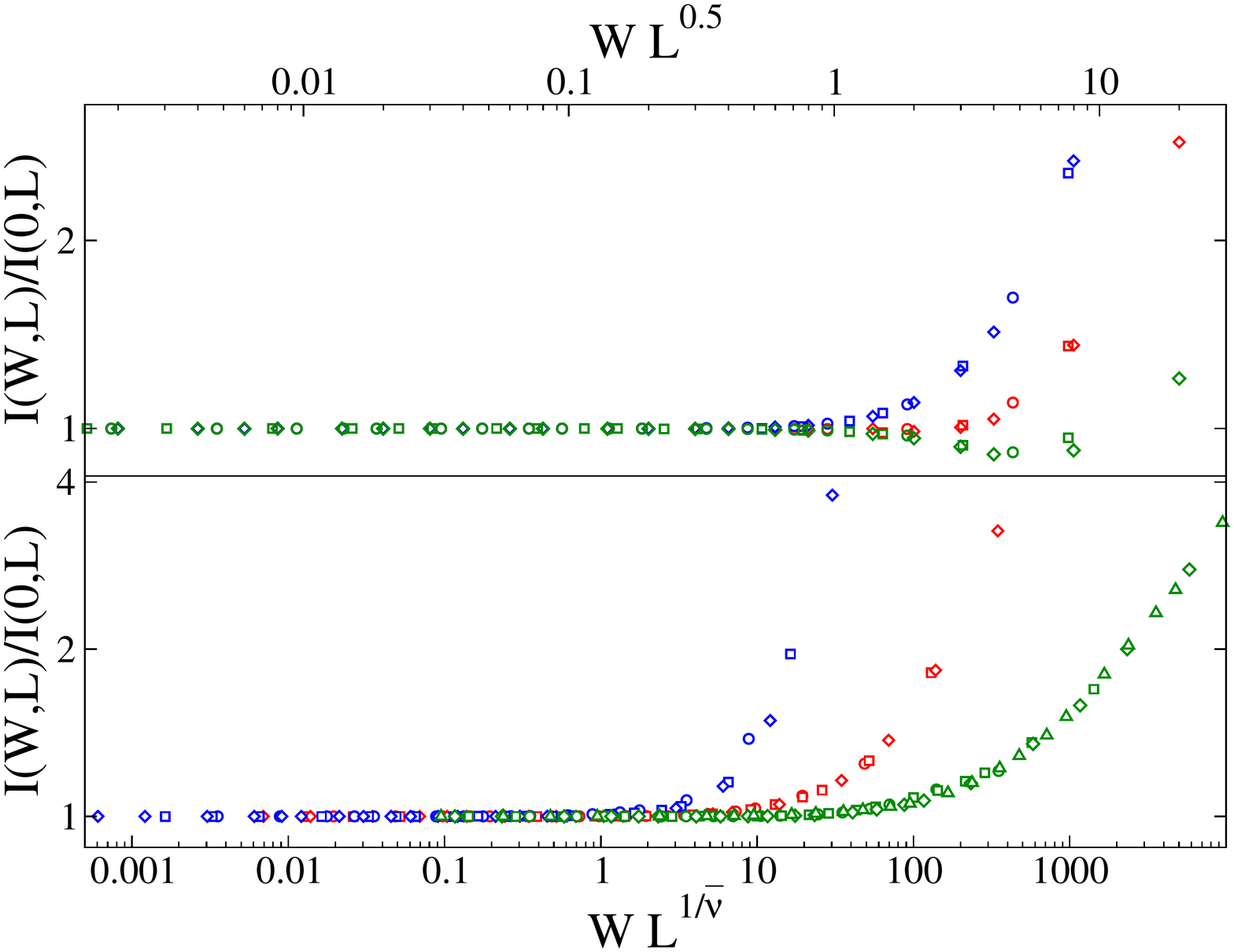}
	\caption{Top panel: Averaged IPR (normalized to ${\cal I}(0,L)$) of the center state ($\alpha=1 + [L_n/2]$) for $n=10$ (circles), $n=12$ (squares), and $n=16$  (diamonds) and for $t_A / t_B=4/5$ (blue), $t_A / t_B=1/2$ (red), and $t_A / t_B=1/3$ (green) showing data collapse as a function of the scaling variable $W L^{0.5}$. Bottom panel: IPR (normalized to ${\cal I}(0,L)$) averaged over $1/16$ of the states around $E=0$ for $n=10$ (circles), $n=12$ (squares), $n=14$  (diamonds), and $n=16$ (up triangles) and for $t_A / t_B=4/5$ (blue), $t_A / t_B=1/2$ (red), and $t_A / t_B=1/3$ (green) showing data collapse as a function of the scaling variable $W L^{1/\overline{\nu}}$.}
\label{center.fig}
\end{figure}

This is highlighted in the top panel of Fig.~\ref{center.fig} which show the scaling plots of the IPR as a function of the disorder $W$ of the state at the center of the spectrum. 
A good collapse of the data for several system sizes is found in terms of the scaling variable $W L^{1/2}$ for all values of $\rho$. Note that a nonmonotonicity of the IPR of the center state starts to appear for $\rho \lesssim 1/2$ and becomes more pronounced as $\rho$ is further decreased.

One can also define quantities {\it averaged over states} in the vicinity of fixed energy or chemical potential. An analysis of finite size scaling of the  IPR at fixed energy $\langle E \rangle$ shows that it is described by a different (nonuniversal) exponent $\overline{\nu}$ for all values of $E$ at fixed $\rho$. 
This is shown, for instance, in the bottom panel of Fig.~\ref{center.fig}, which exhibits the scaling plots of the IPR averaged over a small but finite fraction of the states around zero energy (in practice here we considered 1/16 of states around $\langle E \rangle = 0$). In this case the data for several different sistem sizes show good collapse when plotted in terms of the scaling variable $W L^{1/\overline{\nu}}$.
Note that no signs of the nonmonotonic behavior is seen in this case. 
The dependence of the exponent $\overline{\nu}$ on the ratio of hopping amplitudes is shown in Fig.~\ref{exponent.fig} (red empty circles).
We observe that $\overline{\nu}$ also decreases (possibly logarithmically) to zero as $\rho \to 0$ and approaches the standard value $\overline{\nu} = 2$ in the pure case, $\rho \to 1$. The same scaling exponent is found for other values of the energy $\langle E \rangle$ in the bulk of the spectrum (away from big gaps).
This analysis underscores the importance of distinguishing between the different situations when analyzing a given experimental system.

\section{RG for pure and for weakly disordered chains} \label{sec:rg}
\begin{figure}[h]
\centering
\hspace{-3.4cm} \includegraphics[scale=0.4]{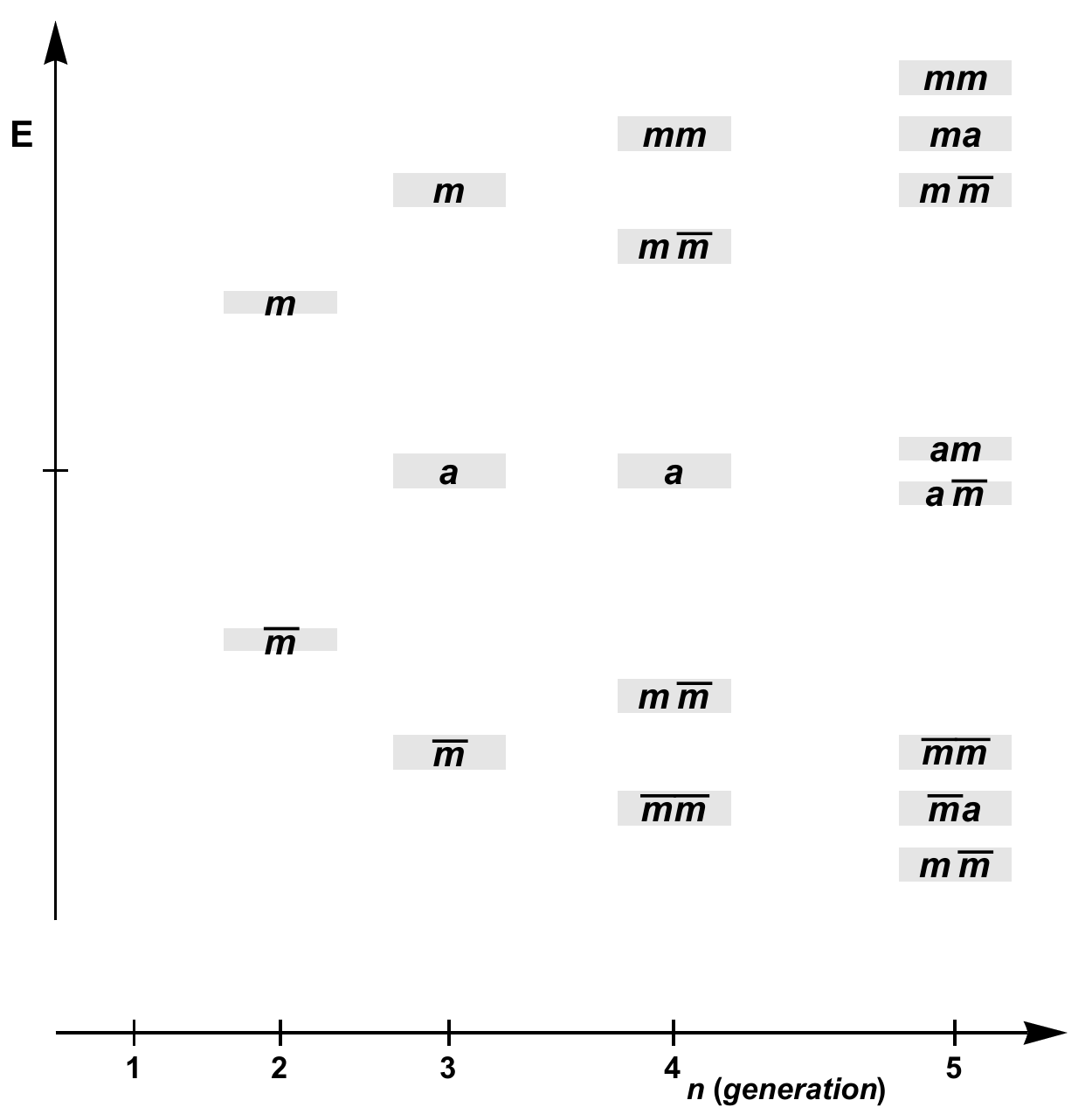} 

\vspace{-5.2cm} \hspace{5cm} \includegraphics[scale=0.5]{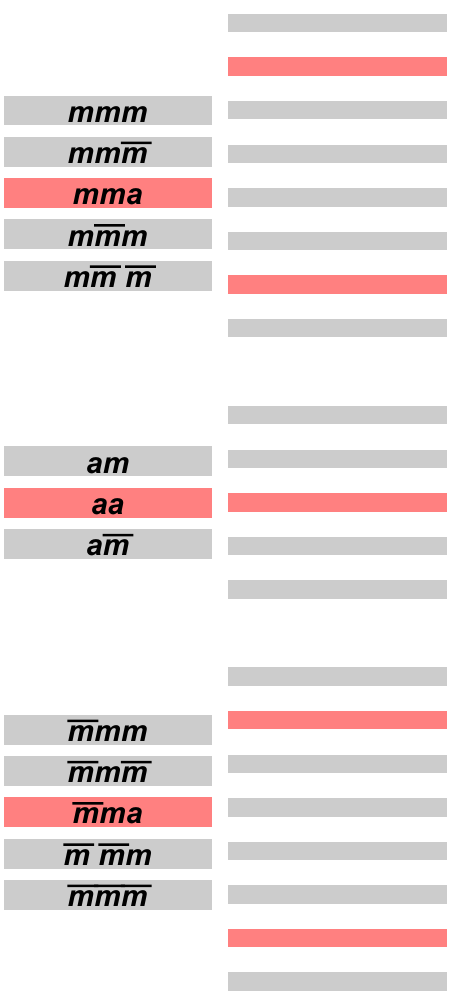} 

\vspace{0.4cm}
\caption {(a) Spectra (in arbitrary units) for successive generations of approximants, showing the recursive structure in RG. (b) The left figure shows the spectrum of a 13-site chain, on the right, the spectrum of a 21-site chain. Bands are colored according to the last RG step: molecular (gray) or atom (red),}
\vspace{-0.1cm}
\label{spectra.fig}
\end{figure}

We begin by 
briefly recall (for details see~\cite{thiem,macePRB2016}) the steps of the real space RG for the pure system before discussing the addition of randomness. Sites on a given chain are termed either ``molecule'' sites---pairs of sites coupled via $t_B$---or  ``atom'' sites---those with  $t_A$ on both sides. 
One defines two different real space decimation procedures:  i) decimating all atoms leaving only sites corresponding to molecules (mRG) or ii)  decimating all molecules leaving only the atom sites (aRG). To lowest nontrivial order in $\rho$ one finds that: under mRG, an initial chain $C_n$ transforms to the chain $C_{n-2}$, with new weaker effective hopping amplitudes $ t'_A$ and $ t'_B$. An energy shift of $\pm t_B$ (resp. $-t_B$) occurs for bonding(m) and antibonding ($\overline{m}$ levels. Under aRG, an initial chain $C_n$ transforms to the chain $C_{n-3}$, with new effective hopping amplitudes given by $t''_A$ and $ t''_B$. The ratio of the strong and weak hopping amplitudes is left invariant in both types of RG. As a result the spectrum of the $n$-th chain can be built up from the spectra of the $n-2$ and $n-3$ generation chains. 
For each energy level $E_\alpha$ ($\alpha=1, \ldots ,L_n$), one can define the ``renormalization path'' or set of characters $a,m,\overline{m},...$. Each element of this RG path is determined by whether the corresponding RG step was atomic or molecular.  $n_m$ denotes the total number of molecular RG steps, and $n_a$ the overall number of atomic RG steps in the RG path. Fig.~\ref{spectra.fig}a) shows schematically how the spectra of chains $n=4,5$ are recursively obtained from spectra of smaller chains, along with the RG paths of levels. Fig..~\ref{spectra.fig}b) shows the spectra of two longer chains, with bands colored according to the last RG transformation (gray for $m(\overline{m})$ and red for $a$), in view of the discussion of the IPRs which will follow. Note that states having the same RG path terminations are expected to have similar properties on large length scales.

To each level described by some RG path, corresponds a wavefunction with support on sites having the {\it{same}} transformation properties under the RG. A wavefunction for energy $E$ for a given chain can be related to the wavefunction of a state  of energy $E'$ on a smaller chain. The scale factors corresponding to mRG and aRG are denoted by $\lambda$ and $\lamb$ respectively, with  
$\vert\psi^{(n)}(i,E)\vert^2 = \lambda \vert\psi^{(n-2)}(i',E')\vert^2$ and $\vert \psi^{(n)}(i,E) \vert^2 =\overline{\lambda} \vert \psi^{(n-3)}(i',E')\vert^2 $,
 where  $i$ and $i'$ correspond to the site indices in the initial and final chains. The existence of two distinct scale factors  
$\lambda$ and $\lamb$ (functions of $\rho$~\cite{macePRB2016}) leads to multifractality of the wavefunctions. 
These recursions relations imply that $\mathcal{I}(E,L_n) \sim L_n^{-D_2(E)}$ where the  exponent $D_2(E)$ which measures the ``mass scaling'' of the atoms associated with the state of energy $E$ depends on its RG path. An explicit calculation
of $D_2(E)$~\cite{thiemschreiberJPCM2011,macePRB2016} shows that, in the limit $n\rightarrow \infty$, states at the edge of the spectrum (for which $n_m=n/2$) are ``more extended'',  i.e. have a larger value of $D_2$, than the state in the center (for which $n_m=0$). 

\subsection{Degenerate perturbation theory for disordered model}
We now extend this RG scheme to our disordered model, for finite chains. We require that $W$ be smaller than the smallest gaps of the spectrum ($W < \zb^{n/3}$). This ensures that the branching hierarchical structure of the spectrum is conserved and the RG path structure of the pure system is not changed (no level crossing due to the random perturbation occurs).  
We aim to compute the renormalized hopping amplitudes, $t'_{A}$ and  $t'_{B}$ obtained for a disordered Fibonacci chain in which the bonds have values  of either $t_A+\epsilon_j$ (weak bond) or  $t_A+\epsilon_j$ (strong bond). Although the onsite energies in the model are taken to be 0, diagonal terms will be generated under RG, and are denoted $\xi_j$, and we will also calculate their renormalized values.  The zero-order Hamiltonian $H_0$ is off-diagonal, and consists only the pure strong couplings $t_B$, while the perturbation $H_1$ contains the weak bonds and diagonal onsite energy terms, $\xi_j$. In Brillouin-Wigner perturbation theory for degenerate states, the effective Hamiltonian is given by~\cite{niunoriPRB}
\begin{equation}
H_{\rm eff} = QH_0Q + QH_1Q + QH_1P\frac{1}{E-H_0}PH_1Q + \ldots
\label{bwpert.eq}
\end{equation}
in the subspace of energy $E$, where the operator $Q=\sum_\alpha \vert\psi_\alpha \rangle\langle\psi_\alpha \vert$ is the projection operator for states in this subspace, and $P=1-Q$.

\begin{figure}[h]
\centering
  \includegraphics[scale=0.6]{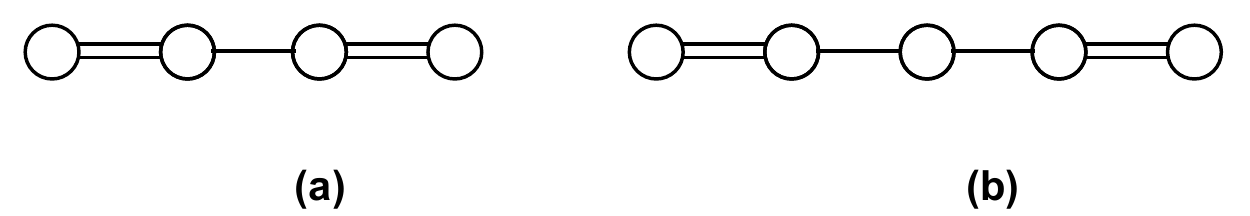}    
\caption {a) Cluster for strong bond calculation (mRG); b) Cluster for weak bond calculation (mRG). }
\label{suppfig1.fig}
\end{figure}

We now illustrate the calculation of parameters of the effective Hamiltonian after a moleculer RG (mRG), namely the onsite energies $\xi'$, and the renormalized strong and weak couplings $t'_B$, $t'_A$. Fig.~\ref{suppfig1.fig}a) shows the cluster of sites which renormalize to give a strong bond after mRG. The three bonds are $t_B+\epsilon_1$,  $t_A+\epsilon_2$ and  $t_B+\epsilon_3$. The onsite energies are zero in the first RG step, but subsequently acquire nonzero values which are denoted by $\xi_j$ ($j=1$ to 3).
The eigenstates of $H_0$ are $\vert\psi_1\rangle=(\vert 1\rangle+\vert 2\rangle)/\sqrt{2}$,  $\vert\psi_2\rangle=(\vert 3\rangle+\vert 4 \rangle)/\sqrt{2}$, corresponding to $E=t_B$, and $\vert\psi_3\rangle=(\vert 1\rangle-\vert 2\rangle)/\sqrt{2}$, and  $\vert\psi_4\rangle=(\vert 3\rangle-\vert 4\rangle)/\sqrt{2}$, corresponding to $E=-t_B$. 

Using  Eq.~(\ref{bwpert.eq})  the onsite energy for the leftmost molecular bonding state, is, to lowest nonvanishing order 
\begin{equation} \label{renorm.eq1}
\xi'_1=\langle\psi_1\vert H_{\rm eff}\vert \psi_1\rangle = \frac{1}{2} (\xi_1+\xi_2) + \epsilon_1 \, ,
\end{equation}
with a similar result for the onsite energy for the right molecular state. The effective (strong) hopping amplitude between the two bonding molecular states is
\begin{equation} \label{renorm.eq2}
 t'_B \equiv  \langle\psi_1\vert H_{\rm eff}\vert \psi_2\rangle = \frac{1}{2} (t_A + \epsilon_2) \, .
 \end{equation}
The renormalized weak coupling is found by considering the cluster in Fig.~\ref{suppfig1.fig}b) consisting of five sites. The eigenstates of $H_0$ are now $\vert\psi_1\rangle=(\vert 1\rangle+\vert 2\rangle)/\sqrt{2}$,  $\vert\psi_2\rangle=(\vert 4\rangle+\vert 5 \rangle)/\sqrt{2}$, corresponding to $E=t_B$, $\vert\psi_3\rangle=\vert 3\rangle$, $\vert\psi_4\rangle=(\vert 1\rangle-\vert 2\rangle)/\sqrt{2}$, and  $\vert\psi_5\rangle=(\vert 4\rangle-\vert 5\rangle)/\sqrt{2}$, corresponding to $E=-t_B$. 
The effective (weak) hopping amplitude is of second order:
\begin{equation} \label{renorm.eq3}
 t'_A  \equiv  \langle\psi_1\vert H_{\rm eff}\vert \psi_2\rangle  
=  \frac{1}{2t_B} (t_A+\epsilon_2)(t_A+\epsilon_3) \, .
\end{equation}

\begin{figure}[h]
\centering
  \includegraphics[width=0.48\textwidth]{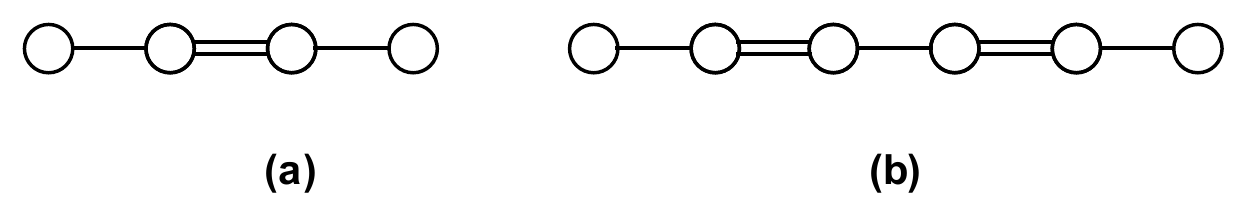}    
\caption {a) Cluster for strong bond calculation (aRG); b) Cluster for weak bond calculation (aRG). }
\label{suppfig2.fig}
\end{figure}

For atomic RG, the clusters to consider for the new strong and weak amplitudes are shown in Fig.~\ref{suppfig2.fig}. The onsite energy and strong and weak hopping amplitudes after aRG  are found to be
\begin{equation} \label{renorm.eq4}
\begin{aligned}
\xi'_1 &= \xi_1 \, , \\
 t'_B &= -(t_A+\epsilon_1)(t_A+\epsilon_3)/t_B \, ,\\
 t'_A &= (t_A+\epsilon_1)(t_A+\epsilon_3)(t_A+\epsilon_5)/t_B^2 \, .
\end{aligned}
\end{equation}

The results of the degenerate perturbation theory, obtained in Eqs.~(\ref{renorm.eq1})-(\ref{renorm.eq4}) in the limit of small $W$, are summarized in the table beow, which gives the onsite energy and the hopping amplitudes up to second order in the perturbations after the first RG step:
\begin{eqnarray}
\begin{array}{|c | c| c|}
\hline 
& mRG & aRG \\
\hline 
\xi_i &  \epsilon & 0 \\
t_B' &  \frac{(t_A+\epsilon_1)}{2}   & - \frac{(t_A+\epsilon)  (t_A+\epsilon')}{t_B} \\
t_A' &  \frac{(t_A+\epsilon) (t_A+\epsilon') }{t_B} & \frac{(t_A+\epsilon) (t_A+\epsilon')  (t_A+\epsilon'')}{t_B^2} \\
\hline
\end{array} 
\label{renorm.eq}
\end{eqnarray}
where as already stated, the $\epsilon$ are i.i.d. random variables. 

Disordering the Fibonacci chain therefore leads to small onsite energy corrections $\xi_i$ (zero for aRG at this order) and modified renormalized hopping amplitudes.  From Eqs.~(\ref{renorm.eq}) one sees that the average renormalized hopping amplitudes are unchanged from their pure values, but their variance is proportional to $W^2$. The spectrum is broadened---i.e. while the average value or center of mass of minibands of the chain are not shifted at lowest order, their widths increase with the disorder strength. For $W$ small enough that levels do not overlap, the RG can therefore proceed as in the pure case. Wavefunctions are determined by the coupling ratio, whose average value is renormalized to $\rho'=\rho +  W^2$ and therefore increases under RG. Since small $\rho$ corresponds to stronger quasiperiodic modulation, one observes that the disorder diminishes the  quasiperiodic modulation---as the number of RG steps increases one gets a homogeneously disordered chain in the large distance limit.

For strong values of disorder gaps are filled in progressively, with the two largest gaps of width $\sim t_B-t_A$ being the last to disappear (when $W$ becomes of the order of $t_B$).  At large disorder, the familiar form of the DOS well-known in the literature of the off-diagonal 1D Anderson model (as reviewed~\cite{andersonmodel}) is recovered. 

\begin{figure}[h]
\centering
\hspace{-0.5cm}\includegraphics[scale=0.27,angle=-90]{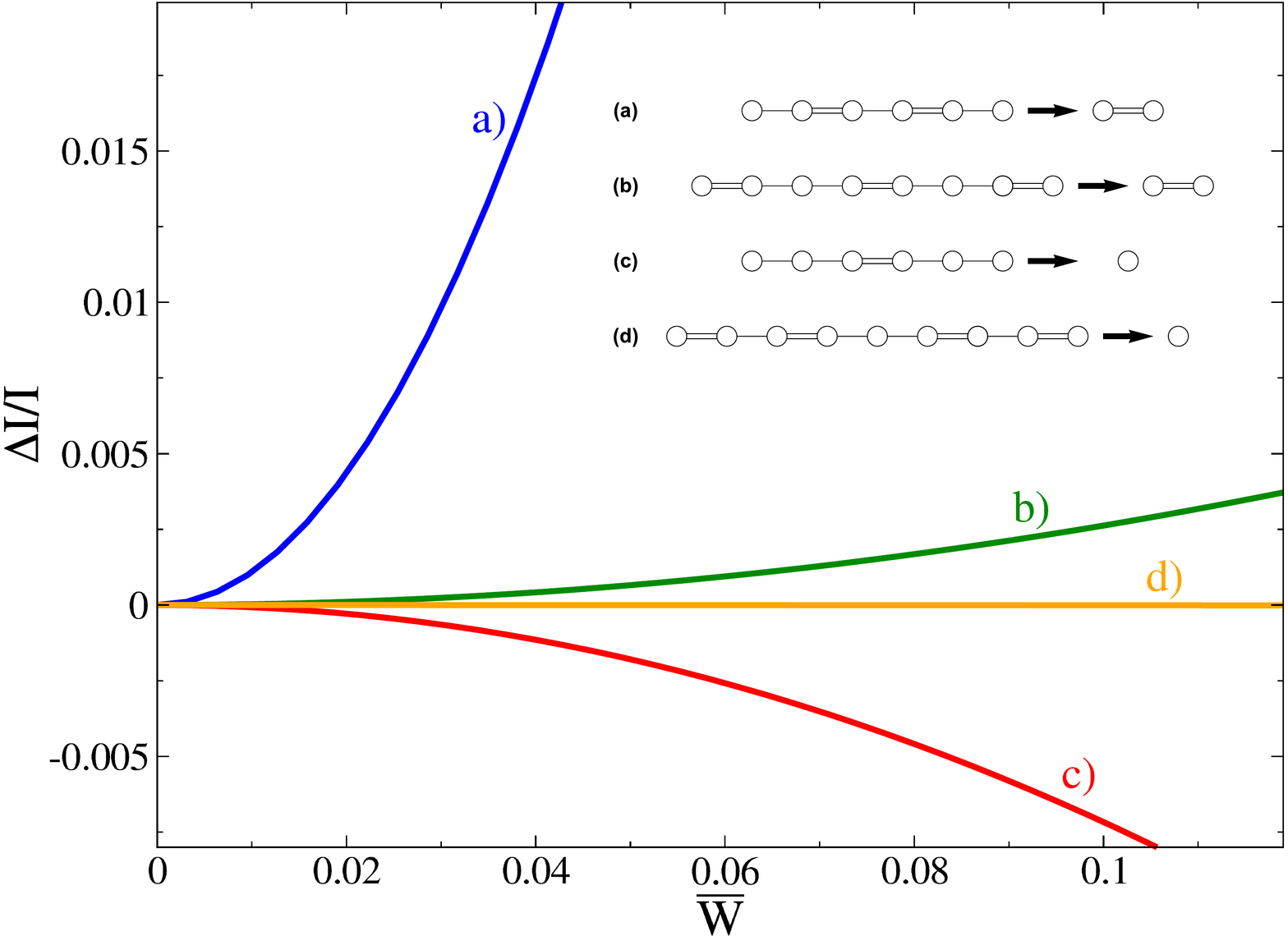} 
\caption {Inset: Clusters transforming to final molecular state a) via mRG and b) via aRG and clusters transforming to final atomic state c) via mRG and d) via aRG. 
Main panel: Plots of $\Delta\mathcal{I}(W)/\mathcal{I}(0)$ for $\rho=0.46$ using the expressions given in Eq.~(\ref{perttheory.eq}).}
\label{clusters.fig}
\vspace{-0.1cm}
\end{figure}

\subsection{IPR corrections due to disorder}
The different types of IPR scaling functions seen in Figs.~\ref{dataIPR.fig} and~\ref{rescaling.fig} 
can now be explained in terms of the nature of the level after the last RG step. As can be seen from Fig.~\ref{spectra.fig}b) the number of ``red'' levels doubles with each mRG so that for large $n$, the number of red levels grows as $2^{n/2}$. The total number of levels grows as $L\sim \omega^{-n}$, so the number of the nonmonotonic ``atom'' states grows with the system size as $L^\beta$, with $\beta = \ln 2/(2 \ln (1/\omega)) \simeq 0.72$. The red levels are those which show a negative IPR change as we will explain below. Considering all combinations of $m$ and $a$ states in the last RG step, we have four possible situations. These four classes of levels have wavefunctions with support on sites which have the same RG characteristics---these are shown in the inset of Fig.~\ref{clusters.fig} along with the final form (molecule or atom). The strong bonds (represented by double lines) of each cluster are taken to be $t_B+\epsilon_j$ and the weak bonds (single lines) are $t_A+\epsilon_j$. The variation of the IPR, $\Delta\mathcal{I}$, can be found by diagonalizing the pure Hamiltonian as a function of $\rho$, computing wavefunction corrections in standard second order perturbation theory in $\epsilon_j$, and finally averaging over all random variables of the cluster.

In order to do this, we consider the hopping model of Eq.~(\ref{eq:H}), where each of the hopping terms is either a perturbed strong  ($t_B+\epsilon_j$) or perturbed weak ($t_A+\epsilon_j$) bond, where $\epsilon_j$ are i.i.d. variables uniformly distributed in $[-W/2,W/2]$. In the pure system, the wavefunctions corresponding to the levels of interest have their support primarily (for small $\rho$) on specific groups of sites arranged as in the four clusters shown in the inset of Fig.~\ref{clusters.fig}. Weak disorder leads to a small redistribution of amplitudes, that we want to compute, perturbatively. We will be interested in the IPR change of specific states:  the band edge molecular level $\alpha=1$ for clusters (a) and (b), and in the atom level close to/at the center for the clusters (c) and (d). The full Hamiltonian is off-diagonal, with $L$ sites and $L-1$ bonds. The latter can be strong or weak bonds, with weak disorder in each of the hopping amplitudes, as given in Eq.~(\ref{eq:H}). The aim of the calculations is to compute the changes of IPR due to the disorder for specific states on each of the clusters, using second order perturbation theory.  

As contrasted with the Brillouin-Wigner perturbation expansion for degenerate states, here we will proceed by splitting the Hamiltonian differently as follows: $H=H_F+H_d$, where the Fibonacci Hamiltonian $H_F$ includes the pure strong and weak bonds ($t_B$ and $t_A$), and $H_d$ contains the disordered part, $\epsilon_j$. 
For a given cluster of $L$ sites, the normalized eigenstates of $H_F$, denoted by $\{\vert\psi_\alpha\rangle\}$ (with $\alpha=1, \ldots ,L$) are nondegenerate and can be computed exactly as a function of $\rho=t_A/t_B$. 
For each of the clusters the IPR of the state $\alpha$ at zero order of the perturbation is then given by $\mathcal{I}(0)=\sum_i \vert \psi_\alpha(i)\vert^4$.

Using standard perturbation theory, the first and second order corrections to the wavefunction are
\begin{displaymath}
\begin{aligned}
 \vert \psi_\alpha\rangle ^{(1)} &= \sum_{\beta \neq \alpha} \frac{\langle \psi_\beta\vert H_d\vert \psi_\alpha\rangle}{E_\alpha-E_\beta} \vert \psi_\beta\rangle \, , \\
 \vert \psi_\alpha\rangle ^{(2)} & = \sum_{\beta \neq \alpha} \bigg[ - \frac{\langle \psi_\alpha\vert H_d\vert \psi_\alpha\rangle \langle \psi_\beta\vert H_d\vert \psi_\alpha\rangle}{(E_\alpha-E_\beta)^2}  \\
 & \qquad \qquad + \sum_{\gamma \neq \alpha} \frac{\langle \psi_\beta\vert H_d\vert \psi_\gamma\rangle \langle \psi_\gamma\vert H_d\vert \psi_\alpha\rangle}{(E_\alpha-E_\beta)(E_\alpha-E_\gamma)} \bigg] \vert \psi_\beta\rangle \, , 
\end{aligned}
\end{displaymath}
which are combined with the zero order term to give $\vert \psi'_\alpha\rangle$. The new IPR is then given by $\mathcal{I}'=\sum_i \vert \psi'_\alpha(i)\vert^4/(\sum_i \vert \psi'_\alpha(i)\vert^2)^2$. As there is no convenient closed form expression as a function of $\rho$ for the fractional change defined by $\Delta {\cal I}/{\cal I}(0) = (\mathcal{I}'-\mathcal{I}(0))/\mathcal{I}(0)$, we will present the results for each of the four states in terms of an expansion around a point $\rho=\rho_0$.
  
\begin{equation}
\begin{aligned}
a) \,\, \frac{ \Delta\mathcal{I}}{\mathcal{I}(0)} \simeq&  \left[
11.32 - 43.83 \, \delta \! \rho  + 147.12 (\delta \! \rho)^2 
\right ] \tilde{W}^2 \, ,\\
b) \,\, \frac{ \Delta\mathcal{I}}{\mathcal{I}(0)} \simeq& \left[
0.15 + 10.98  \, \delta \! \rho - 4.11 (\delta \! \rho)^2
\right] \tilde{W}^2 \, ,\\
c)  \,\, \frac{ \Delta\mathcal{I}}{\mathcal{I}(0)}\simeq&\left[ 
-0.77 + 5.25 \, \delta \! \rho + 10.19 (\delta \! \rho)^2
\right] \tilde{W}^2  \, ,\\
d) \,\,  \frac{ \Delta\mathcal{I}}{\mathcal{I}(0)} \simeq &  \left[
-0.14 + 13.6 \, \delta \! \rho + 3.22 (\delta \! \rho)^2
\right] \tilde{W}^2 \, ,
\end{aligned}
\label{perttheory.eq}
\end{equation}
with $\delta \! \rho = \rho - \rho_0$, for $\rho_0=0.45$. In these expressions, $\tilde{W}$ denotes the width of effective renormalized distribution, which increases with the number of RG steps. Fig.~\ref{clusters.fig} shows plots of these four functions for $\rho = 0.46$. They provide qualitative indications of the behavior of these states since the theory is quantitatively accurate only for values $\rho\ll 1$. 

\begin{figure}[h]
\includegraphics[width=0.48\textwidth]{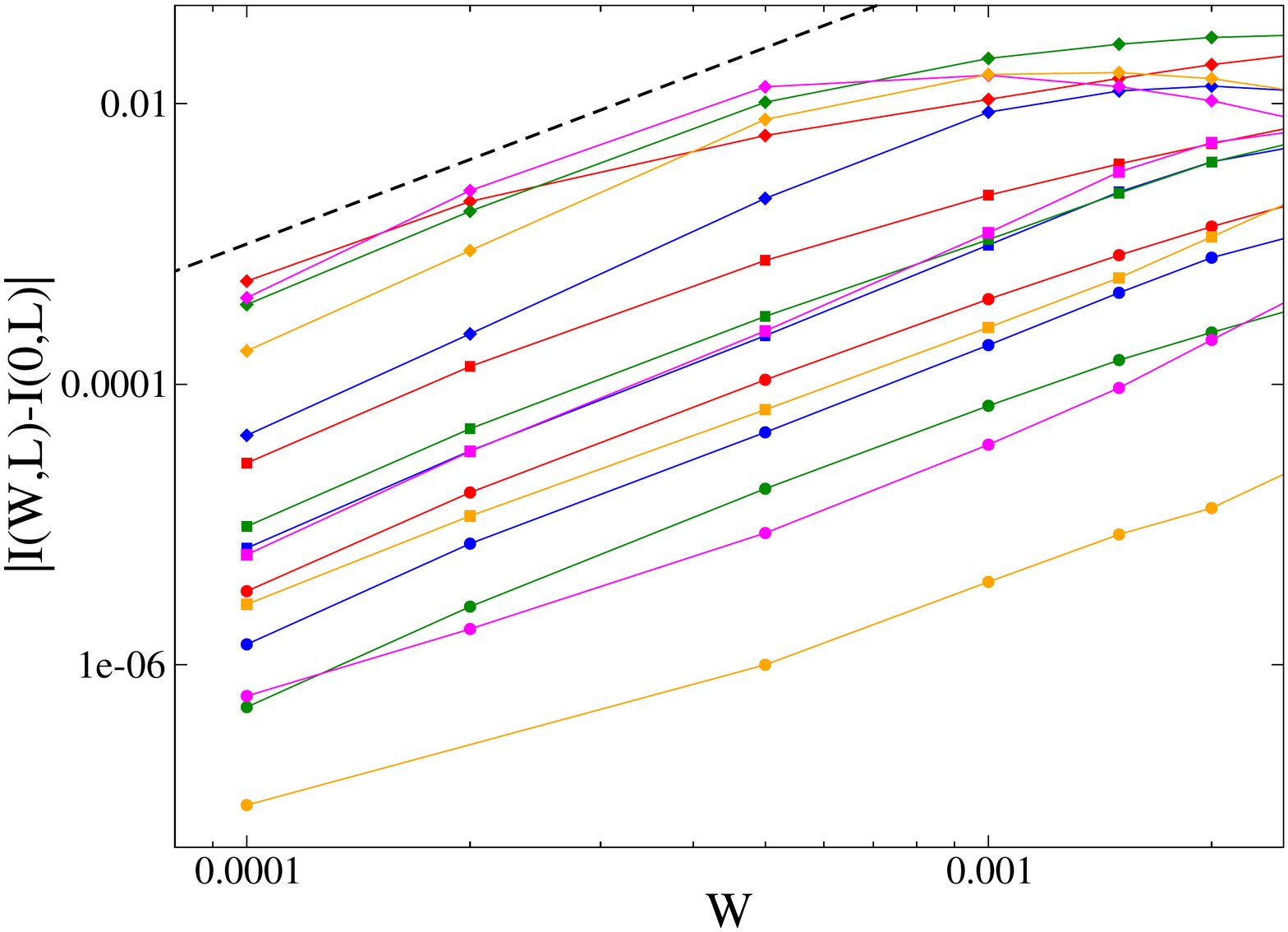}
	\caption {Log-log plot of ${\cal I}_\alpha (W,L) - {\cal I}_\alpha (W,0)$ as a function of the disorder strength $W$ in the $W \to 0$ region for $n=12$ ($L=233$), for several states, $\alpha=1$ (red), $\alpha=3$ (blue), $\alpha=6$ (green), $\alpha=7$ (magenta), and $\alpha=28$ (orange), and for several values of the ratio $t_A/t_B$, $\rho=1/2$ (squares), $\rho=1/3$ (circles), and $\rho=1/5$ (diamond). The dashed line correspond to a quadratic function.} 
\label{quadratic.fig}
\end{figure}

Exact diagonalization results confirm that the IPR of all levels depends quadratically on the disorder for $W \to 0$.
In particular, in Fig.~\ref{quadratic.fig} we show the results of exact diagonalizations for $| {\cal I}_\alpha (W,L) - {\cal I}_\alpha (W,0) |$ as a function of $W$ in the small disorder region for $n=12$ ($L=233$) and for several states (the same as Fig.~\ref{rescaling.fig}) and three different values of $\rho$. This plot confirms that the IPR of all states behaves quadratically for $W \to 0$ as 
\begin{displaymath}
{\cal I}_\alpha (W,L_n) \simeq {\cal I}_\alpha (0,L_n) + c_\alpha^{(n)} \, W^2 \, ,
\end{displaymath}
with positive or negative coefficients $c_\alpha^{(n)}$ depending on the level index (e.g., $c_\alpha^{(n)}<0$ for $\alpha = 3$, $7$, $11$, $15$, $20$, $24$, $28$, $32$, $36$, $\ldots$). This is precisely the behavior predicted by the real space RG approach [see Eq.~(5) of the main text].

Comparing with the $W\rightarrow 0$ regime of Fig.~\ref{dataIPR.fig}, these results for the curvature correspond to the initial behavior of the curves: levels of type (a) and (b) corresponding to molecular final states have positive curvatures, while (c) and (d) corresponding to atomic final states have negative curvatures.  
Case (a) describes band edge states, which localize the most rapidly as disorder is increased compared to the other curves. More differences between levels will appear when longer range structural information is included. These results 
are not qualitatively changed upon adding onsite randomness.

\begin{figure}[h]
\centering
\includegraphics[width=0.48\textwidth]{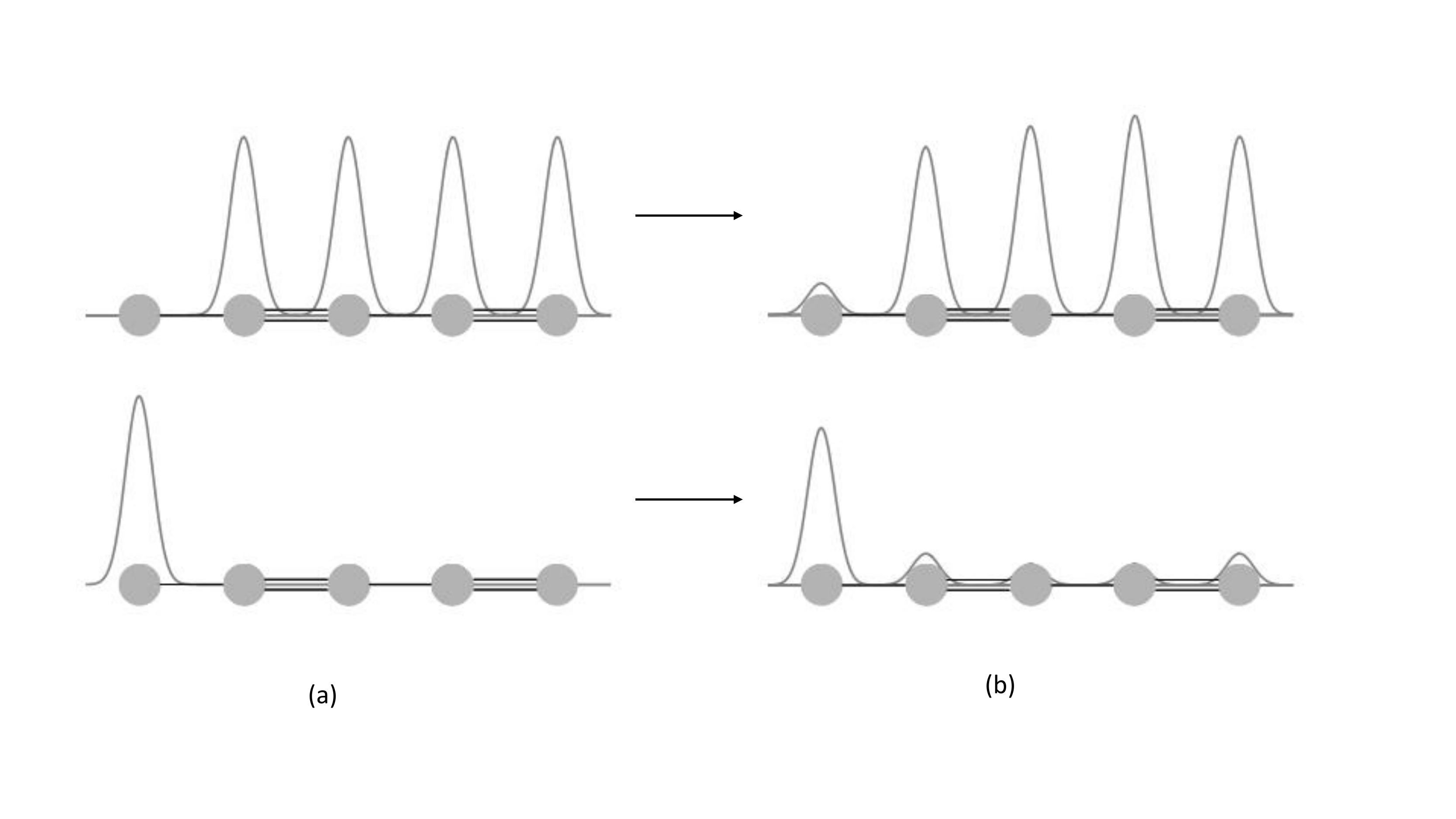} 
\vspace{-1.2cm}
\caption {Schemas for two kinds of wavefunctions (a) molecular wavefunction for pure (continuous) and weakly disordered (dashed) chain.  (b) Atom wavefunctions before and after adding weak disorder. $\Delta$, the change of the IPR, is positive (resp negative) for the two cases.}
\label{newclusters.fig}
\vspace{-0.1cm}
\end{figure}

The sign of the IPR change can be explained in terms of a very simple general argument which is not restricted to the specific case of the Fibonacci chain, as shown schematically in Fig~\ref{newclusters.fig}. When the wavefunction in the pure systems is maximally delocalized, as for the m-state (top figure), disorder tends to {\it increase} the IPR---as seen from the change of the wavefunction. When on the other hand the initial wavefunction is maximally localized (as for the atom state in the lower figure), disorder leads to a {\it decrease} of the IPR. 

To repeat, the behavior shown in Fig.~\ref{clusters.fig} is expected only for very small disorder. For large $W$, differences between ``strong'' and ``weak'' bonds cease to exist and the standard Anderson model is recovered, in which the IPR increases with $W$ (as discussed above, this occurs when the disorder strength becomes of the order of the gaps with the neighboring states and level repulsion sets in). The negative-curvature of $\Delta(\mathcal{I})/\mathcal{I}$ will therefore eventually ``bottom out'' and start increasing with $W$, as we saw in Fig.~\ref{rescaling.fig}. [Notice that the IPR of all the states tend to $1/2$ in the $W \to \infty$ limit for the model described by the Hamiltonian~(\ref{eq:H}).]
  
\section{Conclusions and discussion} \label{sec:conclusions}
To sum up, we have shown, by considering finite approximants of the Fibonacci chain, that the addition of disorder in a model with critical states can lead to an effect of delocalization followed by localization of a subset of levels. We have presented an argument to explain the disorder-dependence of IPR for levels as a function of their RG path. We stress that this phenomenon is not restricted to this specific model~\cite{Huse,sutradhar}, however in the Fibonacci chain one can predict the total number and the energies of such states thanks to the underlying RG scheme. 

The states which exhibit nonmonotonic behavior are those which under RG ultimately are reduced to a single ``atom'' level ($\alpha = 3$, $7$, $11$, $15$, $20$, $24$, $28$, $32$, $36$, $\ldots$), while the other levels show the expected monotonic increase of the IPR with $W$. The number of the nonmonotonic ``atom'' states grows with the system size as $L^\beta$, with $\beta\simeq 0.72$.  
For stronger values of disorder, an upturn of the curves will eventually occur, as the localization lengths become smaller than the chain length, and the usual Anderson localization physics is recovered.  The exponent $\nu$ describing the approach to strong localization depends on the parameter $\rho$, tending to very small values for $\rho\rightarrow 0$, beyond the region of reliability of the numerical computations. The scaling functions for the IPR are different according to the nature of the level (and some of them are nonmonotonic). These characteristics show that this localization transition is in a different universality class from the standard Anderson model, which is recovered in the limit $\rho=1$, with the value $\nu=2/3$ as found previously in a different context~\cite{continentino}. A similar result has been recently found in~\cite{Huse} where it was shown that Anderson localization in a 2D generalization of the Aubry-Andr\'e model appears to be in a quite different universality class from the same model with random potentials.

Some works~\cite{huang,velin} incorporate geometrical forms of disorder
where segments of the chains are flipped. From analyzing Lyapunov exponents using RG~\cite{velin} 
and by direct transfer matrix methods, it was concluded that localization does not occur in this case. More detailed finite size scaling analyses of the phason disordered
1D Fibonacci chains are probably necessary before this issue can be definitively settled. In this context, we note that a similar type of geometrical disorder is considered in a 2D model, and shown to lead to localization of the ground state~\cite{JeanNoel-Julien}. 

The reentrant delocalization-localization of certain states could also be observable in experiments on multilayer systems, by means of precise measurements of the transport in mesoscopic samples. In this context it can be noted that, for three dimensional quasicrystals such as AlCuFe, it was long ago pointed out that structural disorder tends to {\it improve} conductivity~\cite{berger}. 

Our results can be expected to have relevance for the debate on many-body localization due to disorder versus localization due to pseudo-disorder~\cite{dave,alet,MBLlit1,MBLlit2,MBLlit3}.
It was indeed observed that, contrary to naive expectations, adding interactions in quasiperiodic systems does not enhance delocalization, and a MBL transition is observed both in Fibonacci spin chains~\cite{alet} and in fermionic Aubry-Andr\'e models~\cite{dave}. Our results suggest that the transition might have an intermediate regime, where finite size effects can be anomalous. Generally speaking, adding perturbations to the pure noninteracting Hamiltonian could produce non-monotonic or re-entrant behavior.
Questions concerning the critical properties for each case are not just theoretical problems, but are now amenable to experimental verification using cold atoms~\cite{experiments1,experiments2,experiments3,coldatoms1,coldatoms2,coldatoms3}. 

Many important theoretical questions remain open. One
concerns the robustness of our findings with respect to the nature of the quenched disorder. 
As discussed above, it seems reasonable to expect that the addition of i.i.d. onsite disorder to the Hamiltonian does not modify the results as this leaves qualitatively unchanged the RG transformations.
In the same spirit, one could wonder whether the choice of a Gaussian distribution of the random hoppings $\epsilon_i$ might alter the scenario discussed here. Although in the context of Anderson localization taking a Gaussian {\it vs} a box distribution leads essentially to the same physical picture, this is an interesting question, since for unbounded $\epsilon_i$'s level crossing is not forbidden even at infinitesimal disorder and the RG path of neighboring levels might get mixed. Heavy-tailed and/or correlated randomness, instead, are expected to alter significantly the present scenario.
Another very interesting research direction 
is to consider other kinds of random perturbation to the Fibonacci approximants. 
Preliminary results 
indicates that the same kind of nonmonotonic behavior of the IPR is observed for the same states (the atomic final states) as for the disordered case in a model where few weak long-range matrix elements are added the pure Hamiltonian~(\ref{eq:H})---thereby transforming the 1D chain into a sparse random matrix 
with a Fibonacci backbone---suggesting that the way in which individual levels respond to perturbations might be a specific (and robust) feature of their individual critical properties.
Finally, it would also be interesting to investigate the region of the phase diagram $t_A > t_B$ ($\rho>1$) for which much less is known even in the pure limit.

Since this paper was submitted, numerical results have been reported for a different quasiperiodic model -- the Harper model~\cite{sutradhar}, for which the authors report nonmonotonic length dependence of the transport and consequent failure of single parameter scaling. Their results, for 1D in particular,  complement our findings for the Fibonacci model and extend the studies to higher dimensions. It will be interesting to study in detail similarities and differences between the two families of quasiperiodic models. 

\section{Acknowledgments} We are grateful to J.-N. Fuchs, J.-M. Luck, N. Mac\'e, F. Piechon, and J. Vidal for many illuminating discussions. P.J. would like to thank the Idex PALM of University Paris-Saclay for financial support during this project. M.T. is a member of the Institut Universitaire de France.


\end{document}